\documentclass[a4paper,12pt,twoside]{article} 
\usepackage{a4wide}
\usepackage{graphicx}
\usepackage{rotating}
 
\newcommand{\beq}{\begin{equation}}
\newcommand{\eeq}{\end{equation}}
\def\gs{\mathrel{ \rlap{\raise
0.511ex \hbox{$>$}}{\lower 0.511ex \hbox{$\sim$}}}} \def\ls{\mathrel{
\rlap{\raise 0.511ex \hbox{$<$}}{\lower 0.511ex \hbox{$\sim$}}}}

\newcommand{\ba}{\begin{array}{c}}
\newcommand{\baz}{\begin{array}{cc}}
\newcommand{\bad}{\begin{array}{ccc}}
\newcommand{\bea}{\begin{equation} \begin{array}{c}}
\newcommand{\eea}{ \end{array} \end{equation}}
\newcommand{\ea}{\end{array}} 

\newcommand{\dmsol}{\mbox{$\Delta m^2_{\odot}$~}}
\newcommand{\dma}{\mbox{$\Delta m^2_{\rm A}$ }}

\newcommand{\mmin}{\mbox{$ m_{\mbox{}_{\rm{MIN}}} $}}
\def\gtap{\mathrel{ \rlap{\raise 0.511ex \hbox{$>$}}{\lower 0.511ex
   \hbox{$\sim$}}}} 
\def\ltap{\mathrel{ \rlap{\raise 0.511ex
   \hbox{$<$}}{\lower 0.511ex \hbox{$\sim$}}}}
   \newcommand{\deltaatm}{\mbox{$\Delta m^2_{31}$}}
   \newcommand{\deltasol}{\mbox{$ \Delta m^2_{21}$}}
   
   \newcommand{\betabeta}{\mbox{$(\beta \beta)_{0 \nu} $}}

   \newcommand{\meff}{\mbox{$\left| < \! m \! > \right|$}}
   \newcommand{\mefff}{\mbox{$ < \! m  \! > $}}
   
   \newcommand{\hbeta}{$\mbox{}^3 {\rm H}$ $\beta$-decay }
   \newcommand{\eV}{\mbox{$ \ \mathrm{eV}$}}

\newcommand{\pmns}{\mbox{$ U_{\rm PMNS}$}}

\newcommand{\sss}{\sin^2 \theta_{12}}

\newcommand{\dm}{\mbox{$\Delta m_{21}^2$~}}

\begin{document}
%
%
%
\begin{center}
\noindent{\tt \bf 
Theoretical Prospects of Neutrinoless Double Beta Decay
}\vspace{4mm}
\renewcommand{\thefootnote}{\fnsymbol{footnote}}

\noindent{
S.~T.~Petcov$^1$\footnote{Also at: Institute of Nuclear Research and
Nuclear Energy, Bulgarian Academy of Sciences, 1784 Sofia, Bulgaria.}
\footnote{Invited talk given at the Nobel Symposium
(N 129) on Neutrino Physics, August 19 - 24, 2004,
Haga Slott, Enk{\"o}ping, Sweden (to be published in the Proceedings of the
Symposium).}
}\vspace{2mm}

\noindent{\small
$^1$ Scuola Internazionale Superiore di Studi Avanzati 
and INFN,
I-34014 Trieste, Italy
}
\end{center}
\renewcommand{\thefootnote}{\arabic{footnote}}
\setcounter{footnote}{0}

\begin{abstract}
 The compelling experimental evidences for 
oscillations of solar and atmospheric 
neutrinos imply the existence of 
3-neutrino mixing in vacuum.
We briefly review the phenomenology 
of 3-$\nu$ mixing, and the 
current data on the 3-neutrino mixing
parameters. 
The open questions
and the main goals of future 
research in the field of neutrino 
mixing and oscillations are outlined.
The predictions for the effective
Majorana mass $\meff$ in $\betabeta-$decay
in the case of 3-$\nu$ mixing 
and massive Majorana neutrinos are reviewed.
The physics potential of the 
experiments, searching for $\betabeta-$decay 
and having sensitivity to $\meff \gtap 0.01$ eV,
for providing information on the type of 
the neutrino mass spectrum,
on the absolute scale of neutrino masses 
and on the Majorana CP-violation 
phases in the PMNS neutrino 
mixing matrix, is discussed. 

\end{abstract}
\vspace{-0.8cm}
\section{{\large Introduction}}

\vspace{-0.3cm}
\hskip 0.6truecm
 There has been a remarkable  
progress in the studies of neutrino
oscillations in the last several years.
The experiments with solar, 
atmospheric and reactor neutrinos
\cite{sol,SKsolar,SKYoichi04,SNO123,SNO123Art,SKatm98,KamLAND}  
have provided compelling evidences for the 
existence of neutrino oscillations 
driven by nonzero neutrino masses and neutrino mixing.
Evidences for oscillations of neutrinos were
obtained also in the first long baseline
accelerator neutrino experiment K2K~\cite{K2K}.

  The idea of neutrino oscillations 
was formulated in \cite{BPont57}.
It was predicted 
in 1967~\cite{BPont67} that the existence of 
$\nu_e$ oscillations would cause a 
``disappearance'' of solar $\nu_e$ on the way to the Earth.
The hypothesis of solar $\nu_e$
oscillations, which (in one variety or another) 
were considered 
from $\sim$1970 on
as  the most natural 
explanation of the observed ~\cite{sol} solar 
$\nu_e$ deficit (see, e.g., 
refs.~\cite{BiPont78, MSW, BiPet87, SPSchlad97}),
has received a convincing
confirmation from the measurement 
of the solar neutrino flux
through the neutral current reaction on
deuterium by the SNO experiment~\cite{SNO123Art}, 
and by the first results of the KamLAND (KL)
experiment~\cite{KamLAND}.
The combined analysis of the solar neutrino data
obtained by Homestake, SAGE, GALLEX/GNO, 
Super-Kamiokande (SK) and SNO
experiments, and of the 
KL reactor $\bar{\nu}_e$ data
\cite{KamLAND}, established the large mixing 
angle (LMA) MSW oscillations/transitions \cite{MSW}
as the dominant mechanism at the origin 
of the observed solar $\nu_e$ deficit
(see, e.g., \cite{SNO3BCGPR}).
The Kamiokande experiment \cite{Kam96}
provided  the first evidences
for oscillations of atmospheric 
$\nu_{\mu}$ and $\bar{\nu}_{\mu}$,
while the data of the 
SK experiment made the case 
of atmospheric neutrino oscillations convincing  
\cite{SKatm98,SKYoichi04}.
Indications for $\nu$-oscillations
were reported by the LSND 
collaboration \cite{LSND}.

 The latest contributions to these 
magnificent progress are the new 
SK data on the $L/E$-dependence of 
the $\mu$-like  atmospheric 
neutrino events \cite{SKdip04},
$L$ and $E$ being the distance traveled 
by neutrinos and the 
neutrino energy,
and the new  
spectrum data of the 
KL and K2K experiments 
\cite{KL766Ats,K2Knu04}.
For the first time the data
exhibit directly the effects of the 
oscillatory dependence on $L/E$ and $ E$ of 
the probabilities of  
$\nu$-oscillations in vacuum \cite{BP69}.
As a result of these developments, 
the oscillations of solar $\nu_e$,
atmospheric $\nu_{\mu}$ and
$\bar{\nu}_{\mu}$, 
accelerator $\nu_{\mu}$ (at $L\sim$250 km)
and reactor $\bar{\nu}_e$ (at $L\sim$180 km), 
driven by nonzero $\nu$-masses 
and $\nu$-mixing, can be considered as 
practically established.

\vspace{-0.50cm}
\section{{\large The Neutrino Mixing Parameters and $\betabeta-$Decay}}

\vspace{-0.20cm}
   The SK atmospheric neutrino 
and K2K data are best described in 
terms of dominant 2-neutrino
$\nu_{\mu} \rightarrow \nu_{\tau}$ 
($\bar{\nu}_{\mu} \rightarrow \bar{\nu}_{\tau}$)
vacuum oscillations.
The best fit values and the 
99.73\% C.L. allowed ranges of the 
atmospheric neutrino 
oscillation parameters
read \cite{SKYoichi04}: 
\beq 
\label{eq:atmrange}
\ba
|\dma| =2.1\times 10^{-3}~{\rm eV^2},~~
\sin^22\theta_{\rm A} = 1.0~; \\  [0.3cm]
|\dma| = (1.3 - 4.2)\times 10^{-3}~{\rm eV^2},~~
\sin^22\theta_{\rm A} \geq 0.85.
\ea
\eeq
%
\noindent It should be noted that the 
signs of $\dma$ and of 
$\cos2\theta_{\rm A}$, if
$\sin^22\theta_{\rm A} \neq 1.0$, 
cannot be determined using the existing data. 

  Combined 2-neutrino oscillation 
analyses of the solar neutrino and the new KL 
spectrum data show \cite{KL766Ats,BCGPRKL2} that the 
$\nu_{\odot}$-oscillation parameters 
lie in the low-LMA region :
$ \dmsol = (7.9^{+0.6}_{-0.5})\times 10^{-5}~{\rm eV^2}$, 
$\tan^2 \theta_\odot = (0.40^{+0.09}_{-0.07})$.
%
\noindent The high-LMA solution (see, e.g., \cite{SNO3BCGPR})
is excluded at $\sim 3.3\sigma$.
Maximal $\nu_{\odot}$-mixing
is ruled out at $\sim 6\sigma$;
at 95\% C.L. one finds
$\cos 2\theta_\odot \geq 0.28$ \cite{BCGPRKL2},
which has important implications 
(see further). One also has: $\dmsol/|\dma| \sim 0.04 \ll 1$. 
 
  The evidences for $\nu$-oscillations 
obtained in the solar and atmospheric neutrino
and KL and K2K experiments imply  
the existence of 3-$\nu$ mixing
in the weak charged lepton current:
\vspace{-0.2cm}
\begin{equation}
\nu_{l \mathrm{L}}  = \sum_{j=1}^{3} U_{l j} \, \nu_{j \mathrm{L}},~~
l  = e,\mu,\tau,
\label{3numix}
\end{equation}
%
\noindent where 
$\nu_{lL}$ are the flavour neutrino fields,
$\nu_{j \mathrm{L}}$ is the 
field of neutrino $\nu_j$ having 
a mass $m_j$ and
$U$ is the Pontecorvo-Maki-Nakagawa-Sakata (PMNS) 
mixing matrix \cite{BPont57,MNS62}, 
$U \equiv \pmns$.
All existing  $\nu$-oscillation 
data, except the data of LSND experiment  
\cite{LSND},
can be described assuming 
3-$\nu$ mixing in vacuum
and we will consider 
this possibility in what follows
\footnote{In the LSND experiment indications 
for $\bar \nu_{\mu}\to\bar \nu_{e}$ oscillations
with $(\Delta m^{2})_{\rm{LSND}}\simeq 
1~\rm{eV}^{2}$ were obtained.
The minimal 4-$\nu$ 
mixing scheme which could 
incorporate the LSND indications for 
$\bar \nu_{\mu}$ oscillations 
is strongly disfavored 
by the data \cite{Maltoni4nu}.
The $\nu$-oscillation 
explanation of the LSND results
is possible assuming 
5-$\nu$ mixing \cite{JConrad}.
The LSND results are being tested 
in the MiniBooNE experiment
\cite{JConrad}.}.

 The PMNS matrix  can be 
parametrized by 3 angles, 
and, depending on whether the massive 
neutrinos $\nu_j$ are 
Dirac or Majorana particles,
by 1 or 3 CP-violation ($CPV$) phases 
\cite{BHP80,SchValle80D81}.
In the standardly used parameterization 
(see, e.g., \cite{BPP1}),
$\pmns$ has the form:
\bea \label{eq:Upara}
\pmns = \left( \bad 
 c_{12} c_{13} & s_{12} c_{13} & s_{13}  \\[0.2cm] 
 -s_{12} c_{23} - c_{12} s_{23} s_{13} e^{i \delta} 
 & c_{12} c_{23} - s_{12} s_{23} s_{13} e^{i \delta} 
 & s_{23} c_{13} e^{i \delta} \\[0.2cm] 
 s_{12} s_{23} - c_{12} c_{23} s_{13} e^{i \delta} & 
 - c_{12} s_{23} - s_{12} c_{23} s_{13} e^{i \delta} 
 & c_{23} c_{13} e^{i \delta} \\ 
                \ea   \right) 
  {\rm diag}(1, e^{i \frac{\alpha_{21}}{2}}, e^{i \frac{\alpha_{31}}{2}}) \, ,
\eea
%
\noindent where 
$c_{ij} = \cos\theta_{ij}$, 
$s_{ij} = \sin\theta_{ij}$,
the angles $\theta_{ij} = [0,\pi/2]$,
$\delta = [0,2\pi]$ is the 
Dirac $CPV$ phase and
$\alpha_{21}$, $\alpha_{31}$
are two Majorana 
$CPV$ phases \cite{BHP80,SchValle80D81}. 
One can identify 
$\dmsol = \Delta m^2_{21} > 0$.
In this case $|\dma|=|\Delta m^2_{31}|\cong |\Delta m^2_{32}|
\gg \Delta m^2_{21}$,
$\theta_{12} = \theta_{\odot}$, 
$\theta_{23} = \theta_{\rm A}$.
The angle $\theta_{13}$ is limited by 
the data from the CHOOZ and Palo Verde
experiments~\cite{CHOOZPV}. 
The presently existing atmospheric neutrino data
is essentially insensitive 
to $\theta_{13}$ satisfying the 
CHOOZ limit \cite{SKYoichi04}.
The probabilities of survival of 
solar $\nu_e$ and reactor $\bar{\nu}_e$,
relevant for the interpretation of
the solar neutrino, KL and CHOOZ data, depend 
in the case of interest, 
$|\Delta m^2_{31}|\gg \Delta m^2_{21}$,
on $\theta_{13}$:\\

$P^{3\nu}_{\rm KL} \cong \sin^4\theta_{13} + 
\cos^4\theta_{13}\left [1 - \sin^2 2\theta_{12}
\sin^2\frac{\Delta m^2_{21}{\rm L}}{4E} \right ]$,~ 
$P^{3\nu}_{\rm CHOOZ} \cong 1 -  
\sin^2 2\theta_{13}
\sin^2\frac{\Delta m^2_{\rm 31}{\rm L}}{\rm 4E}$,\\

$P^{3\nu}_{\odot} \cong \sin^4\theta_{13} + 
\cos^4\theta_{13}~P^{2\nu}_{\odot}
(\Delta m^2_{21},\theta_{12};N_e\cos^2\theta_{13})$,\\

\noindent where $P_{\odot}^{2\nu}$ 
is the solar $\nu_e$
survival probability \cite{SP88} 
corresponding to 2-$\nu$ oscillations
driven by $\Delta m^2_{21}$ and $\theta_{12}$,
in which the solar $e^-$ number density $N_e$
is replaced by $N_e \cos^2\theta_{13}$ \cite{3nuSP88},
$P^{2\nu}_{\odot} = \bar{P}^{2\nu}_{\odot} + P^{2\nu}_{\odot~{\rm osc}}$,
$P^{2\nu}_{\odot~{\rm osc}}$ being 
an oscillating term 
\cite{SP88} and\\
\begin{equation}
\ba
\bar{P}^{2\nu}_{\odot} = \frac {1}{2} + 
 (\frac{1}{2} - P')\cos2\theta_{12}^m(t_0)\cos 2\theta_{12}, \\ [0.3cm]
 P'= {{e^{-2\pi r_{0}{\Delta m^2_{21}\over {2E}}\sin^2\theta_{12}}
          - e^{-2\pi r_{0}{\Delta m^2_{21}\over {2E}}}}
       \over {1 - e^{-2\pi r_{0}{\Delta m^2_{21}\over {2E}}}}}. 
\ea
\label{Psolexp}
\end{equation}
%
\noindent  Here $\bar{P}^{2\nu}_{\odot}$
is the average probability \cite{Parke86,SP88},
$P'$ is the ``double exponential'' 
jump probability \cite{SP88} 
and $r_0$ is the ``running''
scale-height of the change of $N_e$ along the 
$\nu$-trajectory in the Sun 
\footnote{
The analyses and the extensive numerical
studies performed in 
\cite{PKSP88,NAOsc99}
show that expression (\ref{Psolexp}) for
$\bar{P}^{2\nu}_{\odot}$ provides
a high precision description 
of the average solar $\nu_e$ 
survival probability in the Sun
for any values of $\Delta m^2_{21}$ and $\theta_{12}$
(the relevant error does not exceed 
$\sim$(2-3)\%), including the values from the LMA region. 
The results obtained recently in \cite{HWLS04}
imply actually that the use of the double 
exponential expression for $P'$ 
for description of the LMA 
transitions brings an 
imprecision in $\bar{P}^{2\nu}_{\odot}$
which does not exceed $\sim 10^{-6}$.
}
\cite{SP88,PKSP88,NAOsc99}.
In the LMA solution region
$P^{2\nu}_{\odot~{\rm osc}} \cong 0$ 
\cite{NAOsc99}.
Using the expressions for 
$P^{3\nu}_{\rm KL}$, $P^{3\nu}_{\rm CHOOZ}$
and $P^{3\nu}_{\odot}$ given above,
the 3$\sigma$ allowed range of 
$|\dma|$ from \cite{SKYoichi04}, 
and performing a combined analysis of
the solar neutrino, CHOOZ and KL data, 
one finds \cite{BCGPRKL2}:
\vspace{-0.1cm}
\beq
\sin^2\theta_{13} < 0.055,~~~~99.73\%~{\rm C.L.}
\label{th13}
\eeq
%

\vspace{-0.2cm}
\noindent Similar constraint is obtained 
from a global 3-$\nu$ oscillation analysis
of solar, atmospheric and
reactor neutrino data 
\cite{Maltoni4nu,3nuGlobal}.
A  combined 3-$\nu$ oscillation analysis 
of the solar neutrino, 
\begin{figure}[htb]
\vskip -0.2cm
\includegraphics[width=14cm,height=8cm]{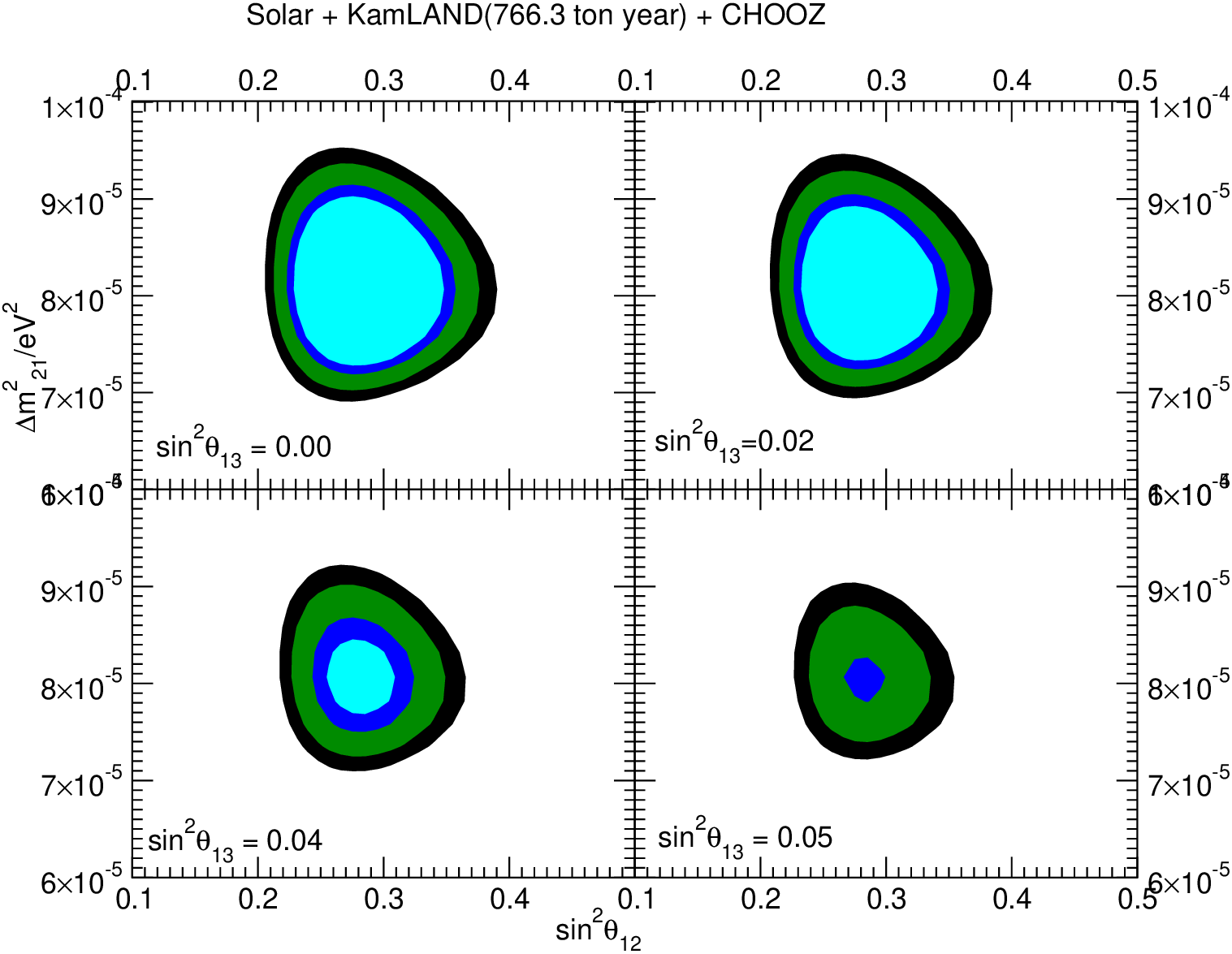} 
\vspace{-0.4cm}
\caption{\label{cont3g}
The 90\%, 95\%, 99\% and 99.73\% C.L. allowed 
regions in the $\dm$-$\sss$ plane, obtained 
in a 3-$\nu$ oscillation analysis of the 
solar neutrino, KL and CHOOZ data
\cite{BCGPRKL2}. 
}
\end{figure}

\noindent  CHOOZ and KL
data shows also \cite{BCGPRKL2}
that for $\sin^2 \theta_{13} \ltap 0.02$
the allowed ranges of 
$\deltasol$ and $\sin^2\theta_{21}$
do not differ substantially 
from those derived in the 
2-$\nu$ oscillation analyzes
(see, e.g., ref.~\cite{KL766Ats}).
The best fit values, 
e.g., read
\footnote{
The best fit value of $\sin^2\theta_{13} = 0.004$  
is different from zero 
but not at statistically significant 
level \cite{BCGPRKL2}.}
\cite{BCGPRKL2}:
\vspace{-0.1cm}
\beq
\deltasol = 8.0\times 10^{-5}~{\rm eV^2},~~~~
\sin^2\theta_{21} = 0.28.
\label{bfvsol}
\eeq
%

\vspace{-0.25cm}
\noindent In Fig. \ref{cont3g} we show the 
allowed regions in the $\dm$-$\sss$ plane,
obtained in a 3-$\nu$ oscillation analysis of the 
solar neutrino, KL and CHOOZ data
for few fixed values of $\sin^2\theta_{13}$.

 As we have seen, the fundamental parameters 
characterizing the 3-neutrino mixing are:
i) the 3 angles $\theta_{12}$, $\theta_{23}$, $\theta_{13}$,
ii) depending on the nature of $\nu_j$ - 1 Dirac ($\delta$), 
or 1 Dirac + 2 Majorana 
($\delta,\alpha_{21},\alpha_{31}$), 
$CPV$ phases, and iii) the 3 neutrino masses, $m_1,~m_2,~m_3$.
It is convenient to express the two 
larger masses in terms of the 
third mass and the measured 
$\dmsol$=$\Delta m^2_{21}>$0 and $\dma$.
We have remarked earlier that the 
atmospheric neutrino and K2K data
do not allow one to determine the sign of 
$\dma$. This implies that if we identify
$\dma$ with $\Delta m^2_{31(2)}$
in the case of 3-neutrino mixing, 
one can have $\Delta m^2_{31(2)} > 0$
or $\Delta m^2_{31(2)} < 0$. 
The two possible signs of
$\dma$ correspond to two 
types of $\nu$-mass spectrum:\\
-- {\it with normal hierarchy},
$m_1 < m_2 < m_3$, $\dma$=$\Delta m^2_{31} >0$,
$m_{2(3)}$=$(m_1^2 + \Delta m^2_{21(31)})^{1\over{2}}$, and\\
-- {\it with inverted hierarchy}
\footnote{In the
convention we use (called A),
the neutrino masses are not 
ordered in magnitude
according to their index number:
$\Delta m^2_{31} < 0$ corresponds to
$m_3 < m_1 < m_2$.
We can also always number the
neutrinos with definite mass 
in such a way that \cite{BGKP96}
$m_1 < m_2 < m_3$. In this convention
(called B), we have in the case of inverted hierarchy spectrum: 
$\dmsol$=$\Delta m^2_{32}$, $\dma$=$\Delta m^2_{31}$.
Convention B is used, e.g., in \cite{BPP1,PPSNO2bb}.
},
$m_3 < m_1 < m_2$,
$\dma$=$\Delta m^2_{32}<$0,
$m_{2}$=$(m_3^2 - \Delta m^2_{32})^{1\over{2}}$, etc.

\noindent The neutrino mass spectrum can also be \\
-- {\it Normal Hierarchical (NH)}: 
$m_1 {\small \ll m_2 \ll }m_3$, $m_2{\small \cong }(\dmsol)^
{1\over{2}}\sim$0.009 eV, 
$m_3{\small \cong }|\dma|^{1\over{2}}$; or\\
-- {\it Inverted Hierarchical (IH)}: $m_3 \ll m_1 < m_2$, 
with $m_{1,2} \cong |\dma|^{1\over{2}}\sim$0.045 eV; or \\
-- {\it Quasi-Degenerate (QD)}: $m_1 \cong m_2 \cong m_3 \cong m_0$,
$m_j^2 \gg |\dma|$, $m_0 \gtap 0.20$ eV. 

   Neutrino oscillation experiments 
allow to determine
differences of squares of neutrino masses,
but not the absolute values of
the masses, or $min(m_j)$. 
Information on the absolute 
scale of $\nu$- masses
can be derived in \hbeta experiments 
\cite{Fermi34,MoscowH3,MainzKATRIN}
and from cosmological and astrophysical data 
(see, e.g., ref. \cite{MTegmark}).
The currently existing most stringent upper 
bounds on the $\bar{\nu}_e$ mass 
were obtained in the Troitzk~\cite{MoscowH3} 
and Mainz~\cite{MainzKATRIN} 
experiments: 
\vspace{-0.1cm}
\begin{eqnarray}
m_{\bar{\nu}_e}  <  2.3 \eV ~~~(95\%~C.L.).
\label{H3beta}
\end{eqnarray}
%

\vspace{-0.1cm}
\noindent We have $m_{\bar{\nu}_e} \cong m_{1,2,3}$
in the case of $QD$ 
$\nu$-mass spectrum.
The KATRIN 
experiment \cite{MainzKATRIN}
is planned to reach a sensitivity  to  
$m_{\bar{\nu}_e} \sim 0.20$ eV,
i.e., to probe the region of the $QD$ 
spectrum. The CMB data of the 
WMAP experiment were used to obtain 
the upper limit \cite{WMAPnu}:
\vspace{-0.1cm}
\begin{eqnarray}
\sum_{j} m_{j} < (0.7 - 2.0)~{\rm eV~~~~~~(95\%~C.L.)},
\label{WMAP}
\end{eqnarray}
%

\vspace{-0.4cm}
\noindent where we have included a conservative 
estimate of the uncertainty in the upper limit 
(see, e.g., \cite{MTegmark}).
The WMAP and future PLANCK 
experiments can be sensitive to 
$\sum_{j} m_{j}  \cong  0.4$ eV.
Data on weak lensing of 
galaxies by large scale structure,
combined with data from the WMAP and PLANCK 
experiments may allow one to determine 
$(m_1 + m_2 + m_3)$ with an uncertainty of 
$\delta \sim (0.04-0.10)$ eV 
(see \cite{MTegmark} and 
the references quoted therein).

  The type of neutrino mass spectrum, i.e., 
$sgn(\deltaatm)$, can be determined 
by studying oscillations of neutrinos and
antineutrinos, say, 
$\nu_{\mu} \leftrightarrow \nu_e$
and $\bar{\nu}_{\mu} \leftrightarrow \bar{\nu}_e$,
in which matter effects are sufficiently large.
This can be done in long base-line 
$\nu$-oscillation experiments 
(see, e.g., \cite{Manfred}).
If $\sin^22\theta_{13}\gtap$0.05
and $\sin^2\theta_{23}\gtap$0.50,
information on $sgn(\Delta m^2_{31})$
might be obtained in 
atmospheric neutrino experiments
by investigating the effects 
of the subdominant transitions
$\nu_{\mu(e)} \rightarrow \nu_{e(\mu)}$
and $\bar{\nu}_{\mu(e)} \rightarrow \bar{\nu}_{e(\mu)}$ 
of atmospheric 
neutrinos which traverse the Earth \cite{JBSP203}.
For $\nu_{\mu(e)}$ ({\it or} $\bar{\nu}_{\mu(e)}$) 
crossing the Earth core, new type of resonance-like
enhancement of the indicated transitions
takes place due to the {\it (Earth) mantle-core
constructive interference effect
(neutrino oscillation length resonance (NOLR))} 
\cite{SP3198}~
\footnote{For the precise conditions
of the mantle-core (NOLR) enhancement
see \cite{SP3198,106107}.
}. 
As a consequence of this effect
\footnote{The Earth mantle-core (NOLR) enhancement of
neutrino transitions differs \cite{SP3198} from the MSW 
one. It also differs \cite{SP3198,106107} from the 
parametric resonance mechanisms of enhancement 
discussed in \cite{Param86}: the conditions
of enhancement found in \cite{Param86}
are not realized for the 
neutrino transitions in the Earth.
In \cite{LiuS98} it is 
erroneously concluded that
the $\nu_{\mu(e)} \leftrightarrow \nu_{e(\mu)}$
transitions of atmospheric neutrinos
crossing the Earth core 
cannot be enhanced by the interplay
of the transitions 
in the Earth mantle and of 
those in the Earth core.
} 
the corresponding 
$\nu_{\mu(e)}$ ({\it or} $\bar{\nu}_{\mu(e)}$)
transition probabilities can be maximal \cite{106107}.
For $\Delta m^2_{31}> 0$, the neutrino transitions
$\nu_{\mu(e)} \rightarrow \nu_{e(\mu)}$
are enhanced, while for $\Delta m^2_{31}< 0$
the enhancement of antineutrino transitions
$\bar{\nu}_{\mu(e)} \rightarrow \bar{\nu}_{e(\mu)}$
takes place, which might allow 
to determine $sgn(\Delta m^2_{31})$.

 After the spectacular experimental 
progress made 
in the studies of $\nu$-oscillations, 
further understanding of the structure of 
neutrino mixing, of its origins  
and of the status of CP-symmetry in 
the lepton sector, requires a large and
challenging program of research to be pursued 
in neutrino physics.  
The main goals of this 
research program should include \cite{SPNu04}:\\
-- High precision measurement of 
neutrino mixing parameters
which control the 
solar and \\
\indent the dominant atmospheric  
neutrino oscillations, 
$\Delta m^2_{21}$, $\theta_{21}$, 
and $\Delta m^2_{31}$, $\theta_{23}$. \\
-- Measurement of, 
or improving by at least a factor 
of (5 - 10) the existing upper 
limit on,\\
\indent $\theta_{13}$ - 
the only small mixing angle  in 
$\pmns$.\\
-- Determination of the
$sgn(\Delta m^2_{31})$ and of
the type of $\nu$-mass spectrum 
($NH,IH,QD$, etc.). \\
-- Determination or 
obtaining significant constraints
on the absolute scale of $\nu$-masses.\\
-- Determination of
the nature--Dirac or Majorana,
of massive neutrinos $\nu_j$. \\ 
-- Establishing whether the CP-symmetry 
is violated in the lepton 
sector a) due to the Dirac \\ 
\indent phase $\delta$, and/or
b) due to the Majorana phases 
$\alpha_{21}$ and $\alpha_{31}$ if 
$\nu_j$ are Majorana particles. \\
-- Searching with increased sensitivity
for possible manifestations, other than 
flavour neutrino \\
\indent oscillations,
of the non-conservation
of the individual lepton charges $L_l$, $l=e,\mu,\tau$, such\\
\indent as $\mu \rightarrow e + \gamma$,
$\tau \rightarrow \mu + \gamma$, etc. decays.\\
-- Understanding at fundamental level 
the mechanism giving rise to  
neutrino masses and mixing and to 
$L_l-$non-conservation. This includes
understanding the origin of the 
patterns of $\nu$-mixing and $\nu$-masses 
suggested by the data. 
Are the observed patterns of 
$\nu$-mixing and of $\Delta m^2_{21,31}$
related to the existence of new  
symmetry of particle interactions?
Is there any relations between quark 
mixing and neutrino mixing?
Is $\theta_{23} = \pi/4$, or $\theta_{23} > \pi/4$
or else $\theta_{23} < \pi/4$?
Is there any correlation between the 
values of $CPV$ phases 
and of mixing angles in  $\pmns$?
Progress in the theory of 
neutrino mixing might also lead, in particular, 
to a better understanding of the 
origin of baryon asymmetry of 
the Universe \cite{LeptoG}.

   The mixing angles, $\theta_{21}$,
$\theta_{23}$ and $\theta_{13}$,
the Dirac $CPV$ phase $\delta$ and 
$\Delta m^2_{21}$ and $\Delta m^2_{31}$ 
can, in principle, be measured with a 
sufficiently high precision in
$\nu$-oscillation experiments 
(see, e.g., \cite{Manfred,th12SKGd}).
These experiments, however,
cannot provide information on the
$\nu$-mass scale
and on the nature of massive neutrinos $\nu_j$,
they are insensitive to
the Majorana $CPV$ phases 
$\alpha_{21,31}$
\cite{BHP80,Lang86}.
Establishing whether $\nu_j$
have distinct antiparticles (Dirac fermions)
or not (Majorana fermions) is of fundamental
importance for understanding 
the underlying symmetries of 
particle interactions 
and the origin of $\nu$-masses. 
If $\nu_j$ are Majorana fermions,
getting experimental information about the  
Majorana $CPV $phases in $\pmns$
would be a remarkably challenging problem
\cite{BGKP96,BPP1,BargerCP,deGBorisRabi,PPR1}. 
The phases $\alpha_{21,31}$ 
can affect significantly 
the predictions for the 
rates of (LFV) decays
$\mu \rightarrow e + \gamma$,
$\tau \rightarrow \mu + \gamma$, etc.
in a large class of supersymmetric theories
with see-saw mechanism of 
$\nu$-mass generation 
(see, e.g., \cite{PPY03}).
Majorana $CPV$ phases might be at 
the origin of baryon asymmetry of 
the Universe \cite{LeptoG}. 

    In the present article
we will review the potential contribution
the studies of neutrinoless double 
beta ($\betabeta-$) decay of even-even
nuclei, $(A,Z) \rightarrow (A,Z + 2) + e^- + e^-$, 
can make to the program of research
outlined above. The \betabeta-decay is allowed 
if the neutrinos with definite mass 
are Majorana particles (for reviews see, 
 e.g., \cite{BiPet87,Morales02,ElliotVogel02,STPFocusNu04}).
Let us recall that the nature - Dirac or Majorana,
of the massive neutrinos 
$\nu_j$, is related to the 
symmetries of particle interactions. Neutrinos $\nu_j$
will be Dirac fermions if the 
particle interactions conserve some 
lepton charge, e.g., the total
lepton charge $L$. If there does not exist 
any conserved lepton charge, neutrinos with 
definite mass can be Majorana particles.
As is well-known, the massive neutrinos are 
predicted to be of Majorana nature
by the see-saw mechanism 
of neutrino mass generation (see, e.g., \cite{seesaw}),
which also provides a  
very attractive explanation of the
smallness of the neutrino masses 
and - through the leptogenesis theory 
\cite{LeptoG},
of the observed baryon asymmetry 
of the Universe.

  If the massive neutrinos $\nu_j$ are Majorana fermions, 
processes in which the total lepton charge
$L$ is not conserved and changes by two units, 
such as $K^+ \rightarrow \pi^- + \mu^+ + \mu^+$,
$\mu^+ + (A,Z) \rightarrow (A,Z+2) + \mu^-$, etc.,
should exist. The process most sensitive to the 
possible Majorana nature of 
neutrinos $\nu_j$ is  
the $\betabeta-$decay
(see, e.g., \cite{BiPet87,APSbb0nu}).
If the \betabeta-decay is generated
{\it only by the (V-A) charged current 
weak interaction via the exchange of the three
Majorana neutrinos}  $\nu_j$ ($m_j \ltap 1$ eV),
which will be assumed in what follows,
the dependence of the 
\betabeta-decay amplitude $A\betabeta$ 
on the neutrino mass and mixing parameters
factorizes in the 
effective Majorana mass $\mefff$
(see, e.g., \cite{BiPet87,ElliotVogel02}):
\vspace{-0.15cm}
\begin{equation}
A\betabeta \sim \mefff~M~,
\label{Abb}
\end{equation}
%

\vspace{-0.3cm}
\noindent where $M$ is the corresponding 
nuclear matrix element (NME) and
\vspace{-0.1cm}
\begin{equation}
\meff  = \left| m_1 |U_{\mathrm{e} 1}|^2 
+ m_2 |U_{\mathrm{e} 2}|^2~e^{i\alpha_{21}}
 + m_3 |U_{\mathrm{e} 3}|^2~e^{i\alpha_{31}} \right|~.
\label{effmass2}
\end{equation}
%

\vspace{-0.1cm}
\noindent
If CP-invariance holds
\footnote{We assume that $m_j > 0$ and that
the fields of the 
Majorana neutrinos $\nu_j$ 
satisfy the Majorana condition:
$C(\bar{\nu}_{j})^{T} = \nu_{j},~j=1,2,3$,
where $C$ is the charge conjugation matrix.
}, 
one has \cite{LW81}
$\alpha_{21} = k\pi$, $\alpha_{31} = 
k'\pi$, where $k,k'=0,1,2,...$, and
\vspace{-0.15cm}
\begin{equation}
\eta_{21} \equiv e^{i\alpha_{21}} = \pm 1,~~~
\eta_{31} \equiv e^{i\alpha_{31}} = \pm 1 
\label{eta2131}
\end{equation}
%

\vspace{-0.2cm}
\noindent represent the relative 
CP-parities of  Majorana neutrinos 
$\nu_1$ and $\nu_2$, and 
$\nu_1$ and $\nu_3$, respectively.
It follows from eq. (\ref{effmass2}) that
the measurement of $\meff$ will provide
information, in particular, on $m_j$.
As eq. (\ref{Abb}) indicates,
the observation of  
\betabeta-decay of a given nucleus and
the measurement of the corresponding
half-life, would allow  to determine 
$\meff$ only if the value of the relevant
NME is known with a relatively small uncertainty.

   The experimental searches for $\betabeta-$decay
have a long history (see, e.g., \cite{Morales02,ElliotVogel02}).
The best sensitivity was achieved in the
Heidelberg-Moscow $^{76}$Ge experiment \cite{76Ge00HM}:
\vspace{-0.2cm}
\begin{equation}
\meff < (0.35 - 1.05)\ \mathrm{eV},~~~90\%~{\rm C.L.}
\label{76Ge00}
\end{equation}
%

\vspace{-0.3cm}
\noindent where a factor of 3 
uncertainty associated with 
the calculation of the relevant
nuclear matrix element 
\cite{ElliotVogel02} is taken 
into account. A positive signal at 
$>$3$\sigma$, corresponding to  
$\meff = (0.1 - 0.9)~{\rm eV}$  at 99.73\% C.L.,
is claimed to be observed in \cite{Klap04}. 
This result will be checked 
in the currently running and future
$\betabeta$-decay experiments. 
However, it may take a long time 
before comprehensive checks 
could be completed.
Two experiments, NEMO3 (with $^{100}$Mo and 
$^{82}$Se) \cite{NEMO3}
and CUORICINO (with $^{130}$Te) \cite{CUORI},
designed to reach sensitivity to 
$\meff\sim$(0.2-0.3) eV, 
are taking data. Their first results read, 
respectively (90\% C.L.):
\vspace{-0.2cm}
\begin{equation}
\meff < (0.7 - 1.2)~\mathrm{eV}~\cite{NEMO3},~~~
\meff < (0.3 - 1.6)~\mathrm{eV}~\cite{CUORI},
\label{NEMO3CUOR}
\end{equation}
%

\vspace{-0.3cm}
\noindent where estimated uncertainties in the NME
are accounted for.
A number of projects aim 
to reach sensitivity to 
$\meff\sim$(0.01--0.05) eV 
\cite{CUORI,APSbb0nu}: 
CUORE ($^{130}$Te), GENIUS ($^{76}$Ge),
EXO ($^{136}$Xe), MAJORANA ($^{76}$Ge),
MOON ($^{100}$Mo), 
XMASS ($^{136}$Xe), etc. 
These experiments
can probe the region 
of $IH$ and $QD$ spectra
and test the positive result 
claimed in \cite{Klap04}. 

 The \betabeta-decay experiments are presently 
the only feasible experiments capable of
establishing the Majorana 
nature of massive neutrinos 
\cite{BiPet87,ElliotVogel02,APSbb0nu}.
As we will discuss in what follows,
a measurement of a nonzero value of 
$\meff \gtap 10^{-2}$ eV:\\
-- Can give also information on 
neutrino mass spectrum 
\cite{PPSNO2bb,PPW,PPRSNO2bb}
(see also \cite{BGGKP99}).\\
-- Can provide unique information on the 
absolute scale of neutrino masses
(see, e.g., \cite{PPW}). \\ 
-- With additional information
from other sources (\hbeta  experiments
and/or cosmological/astrophysical data) 
on the absolute $\nu$-mass scale,
can provide unique information
on the Majorana 
$CPV$ phases $\alpha_{21}$ and/or $\alpha_{31}$ 
\cite{BPP1,BGKP96,PPR1,PPW,WR00}.

\vspace{-0.6cm}
\section{{\large Properties of Majorana Neutrinos: Brief Summary 
}}
\vspace{-0.3cm}

 The properties of Majorana fields (particles) are
very different from those of Dirac fields (particles).
A massive Majorana neutrino $\chi_k$ 
can be described (in local quantum field theory)
by 4-component complex spin 1/2 field $\chi_k(x)$
which satisfies the Majorana condition:
\vspace{-0.1cm}
\beq
C~(\bar{\chi}_k(x))^{{\rm T}} = \xi_k \chi_k(x), ~~|\xi_k|^2 = 1.
\label{MajCon}
\eeq
%

\vspace{-0.2cm}
\noindent where $C$ is the charge conjugation matrix.
The Majorana condition is 
invariant under {\it proper} Lorentz transformations.
It reduces by 2 the number of independent 
components in $\chi_k(x)$.

  The condition (\ref{MajCon}) is invariant with
respect to $U(1)$ global gauge transformations
of the field $\chi_k(x)$ carrying a $U(1)$ charge $Q$,
$\chi_k(x) \rightarrow e^{i\alpha Q}\chi_k(x)$, only if $Q = 0$.
As a result, i) $\chi_k$ cannot carry nonzero 
additive quantum numbers (lepton charge, etc.),
and ii) the field $\chi_k(x)$ cannot ``absorb'' phases
\footnote{
This is the reason why the $PMNS$ matrix 
contains two additional $CPV$ phases
in the case of massive Majorana neutrinos \cite{BHP80}.
}.
Thus, $\chi_k(x)$ describes 2 spin states of a spin 1/2, 
{\it absolutely neutral particle}, which is identical with
its antiparticle, $\chi_k \equiv \bar{\chi}_k$.
If $CP$-invariance holds, Majorana neutrinos 
have definite $CP$-parity $\eta_{CP}(\chi_k) = \pm i$:
\vspace{-0.2cm}
\beq
U_{CP}~\chi_k(x)~U^{-1}_{CP} = \eta_{CP}(\chi_k)~\gamma_{0}~\chi_k(x'),~~
\eta_{CP}(\chi_k) = \pm i~.
\label{CPMaj} 
\eeq
%

\vspace{-0.2cm}
  It follows from the Majorana condition that 
the currents $\bar{\chi}_{k}(x)O^{i}\chi_{k}(x) \equiv 0$,
${\rm for}~O^{i}=\gamma_{\alpha}$;
$\sigma_{\alpha\beta}$; $\sigma_{\alpha\beta}\gamma_5$.
This means that Majorana fermions (neutrinos)
cannot have nonzero $U(1)$ charges and 
intrinsic magnetic and electric dipole moments.

  Finally, if $\Psi(x)$ is a Dirac field 
and we define the standard propagator of $\Psi(x)$ as
\vspace{-0.1cm}
\beq
 <0|T(\Psi_{\alpha}(x)\bar{\Psi}_{\beta}(y))|0> = 
S_{\alpha\beta}^{F}(x-y)~,
\label{DiracP} 
\eeq
%

\vspace{-0.2cm} 
\noindent one has
\beq
<0|T(\Psi_{\alpha}(x)\Psi_{\beta}(y))|0> = 0~,~~
<0|T(\bar{\Psi}_{\alpha}(x)\bar{\Psi}_{\beta}(y))|0> = 0~.
\label{DNSProp}
\eeq
%
\noindent In contrast, a Majorana neutrino 
field $\chi_k(x)$ has, in addition to the standard propagator
\beq
<0|T(\chi_{k\alpha}(x)\bar{\chi}_{k\beta}(y))|0> = 
S_{\alpha\beta}^{Fk}(x-y)~,
\label{MajSP}
\eeq
%

\vspace{-0.2cm}
\noindent two non-trivial {\it non-standard (Majorana)} propagators
\footnote{
For further detailed discussion of the properties of 
Majorana neutrinos (fermions) see, e.g., \cite{BiPet87}.
}
\vspace{-0.1cm}
\beq
\ba
<0|T(\chi_{k\alpha}(x)\chi_{k\beta}(y))|0> = 
-\xi^{*} S_{\alpha\delta}^{Fk}(x-y) C_{\delta\beta}~,\\[0.2cm]
<0|T(\bar{\chi}_{k\alpha}(x)\bar{\chi}_{k\beta}(y))|0> = 
\xi~C^{-1}_{\alpha\delta}S_{\delta\beta}^{Fk}(x-y)~.
\label{MajNSP}
\ea
\eeq
%

\vspace{-0.2cm}
\noindent This result implies that 
if $\nu_j(x)$ in eq. (\ref{3numix}) 
are massive Majorana neutrinos,
$\betabeta$-decay can proceed
by exchange of virtual neutrinos $\nu_j$ since  
$<0|T(\nu_{j\alpha}(x)\nu_{j\beta}(y))|0> \neq 0$.

\vspace{-0.6cm}
\section{{\large Predictions for the Effective Majorana Mass 
}}
\vspace{-0.2cm}

The predicted value 
of \meff{} depends in the 
case of $3-\nu$ mixing on~\cite{SPAS94}
(see also \cite{BPP1,BGGKP99}):
i) $\dma = \Delta m^2_{31(32)}$,
ii) $\theta_{\odot} = \theta_{12}$ and 
$\Delta m^2_{\odot}= \Delta m^2_{21}$, 
iii) the lightest neutrino mass, $min(m_j)$
and on iv) the mixing angle $\theta_{13}$.
In the convention (A) employed by us, 
one has
$|U_{\mathrm{e} 1}|^2 = \cos^2\theta_{\odot} (1 - |U_{\mathrm{e} 3}|^2)$, 
$|U_{\mathrm{e} 2}|^2 = \sin^2\theta_{\odot} (1 - |U_{\mathrm{e} 3}|^2)$,
and  $|U_{\mathrm{e} 3}|^2 \equiv \sin^2\theta_{13}$.

  Given $\dmsol$, $\dma$, $\theta_{\odot}$ and
$\sin^2\theta_{13}$, the value of $\meff$ 
depends strongly on the type of the
neutrino mass spectrum as well as 
on the values of the two
Majorana $CPV$ phases
of the PMNS matrix,
$\alpha_{21,31}$ 
(see eq.\ (\ref{effmass2})).
In the case of QD spectrum
($m_1 \cong m_2 \cong m_3 = m_0$, 
$m_{0}^2 \gg |\dma|,\deltasol$),
$\meff$ is essentially independent on
$\dma$ and $\dmsol$, and 
the two possibilities, 
$\dma > 0$ and 
$\dma < 0$, lead {\it effectively} 
to the same predictions for $\meff$.

\vspace{0.2cm}
{\bf Normal Hierarchical Neutrino Mass Spectrum.}
In this case one has \cite{BPP1}
\vspace{-0.1cm}
\begin{equation}
\meff = \left|(m_1 \cos^2\theta_\odot  + 
e^{i\alpha_{21}} \sqrt{\dmsol}
\sin^2 \theta_\odot)\cos^2\theta_{13}\right.
+ \left. 
\sqrt{\dma} \sin^2\theta_{13}~e^{i\alpha_{31}} \right| 
\label{meffNH1}
\end{equation}
\vspace{-0.8cm}
\begin{equation}
\simeq \left| \sqrt{\dmsol}
\sin^2 \theta_\odot \cos^2\theta_{13} + 
\sqrt{\dma} 
\sin^2\theta_{13} e^{i(\alpha_{31} - \alpha_{21})} \right| \nonumber
\label{meffNH2}
\end{equation}
%

\vspace{-0.1cm}
\noindent where we have neglected the 
term  $\sim m_1$ in eq. (\ref{meffNH2}).
Although  neutrino $\nu_1$
effectively ``decouples'' and does not 
contribute to $\meff$, eq. (\ref{meffNH2}),  
the value of $\meff$ 
still depends on the Majorana $CPV$
phase $\alpha_{32} = \alpha_{31} - \alpha_{21}$.
This reflects the fact that in contrast
to the case of massive Dirac 
neutrinos (or quarks),
$CP$-violation can take place
in the mixing of only two massive 
Majorana neutrinos \cite{BHP80}.   
Further, since \cite{BCGPRKL2} 
$\sqrt{\dmsol} \ltap 9.5\times 10^{-3}$ eV,
$\sin^2\theta_{\odot} \ltap 0.36$,
$\sqrt{\dma} \ltap 5.4\times 10^{-2}$ eV 
\cite{Maltoni4nu} (at 90\% C.L.),
and the largest neutrino mass
enters into the expression for $\meff$ with
the factor $\sin^2\theta_{13} < 0.055$, 
the predicted value of $\meff$ is 
typically $\sim few\times 10^{-3}$ eV: 
for $\sin^2\theta_{13} = 0.04~(0.02)$ one finds
$\meff \ltap 0.005~(0.004)$ eV.
Using the best fit values of the indicated 
parameters (see eqs. (\ref{eq:atmrange}) 
and (\ref{bfvsol})) we get 
$\meff \ltap 0.0044~(0.0035)$ eV.
It follows from eq. (\ref{meffNH1}) 
and the allowed ranges of values of 
$\dmsol$, $\dma$,
$\sin^2\theta_{\odot}$,
$\sin^2\theta_{13}$ as well as of 
the lightest neutrino mass $m_1$ and
$CPV$ phases $\alpha_{21,31}$ 
that in the case of spectrum
with {\it normal hierarchy} 
there can be a complete 
cancellation between 
the three terms in eq. (\ref{meffNH1}) and
one can have \cite{PPW} $\meff = 0$.

\vspace{0.2cm} 
{\bf Inverted Hierarchical Spectrum.}
For $IH$ neutrino mass spectrum 
(see, e.g., \cite{BPP1}) 
$m_3 \ll m_1 \cong m_2 \cong \sqrt{|\dma|} = \sqrt{\Delta m^2_{23}}$. 
Neglecting $m_3 \sin^2\theta_{13}$ in eq. (\ref{effmass2}),
we find \cite{BGKP96}:
\vspace{-0.1cm}
\begin{equation}
\meff \cong \sqrt{|\dma|} \cos^2\theta_{13} \sqrt{1 - \sin^22\theta_{\odot} 
\sin^2\frac{\alpha_{21}}{2}}.
\label{meffIH1}
\end{equation}
%

\vspace{-0.2cm}
\noindent Even though one of the
three massive Majorana neutrinos 
``decouples'', the value of $\meff$
depends on the Majorana CP-violating phase
$\alpha_{21}$. Obviously, 
\vspace{-0.1cm}
\begin{equation}
\sqrt{|\dma|}~|\cos 2 \theta_\odot|~\cos^2\theta_{13}
\leq~ \meff \leq \sqrt{|\dma|}\cos^2\theta_{13}.
\label{meffIH2}
\end{equation}
%

\vspace{-0.1cm}
\noindent  The upper and the lower limits
correspond respectively to the 
$CP$-conserving cases. 
Most remarkably,
since according to the 
solar neutrino and KamLAND data
$\cos 2 \theta_\odot \sim 0.40$,
we get a significant lower limit on 
$\meff$, typically exceeding $10^{-2}$ eV,
in this case \cite{PPSNO2bb,PPW}.
Using, e.g., the best 
fit values of $|\dma|$ and 
$\sin^2\theta_{\odot}$ one finds:
$\meff \gtap 0.02$ eV.
The maximal value of $\meff$ is determined 
by $|\dma|$ and can reach, as it follows from
eqs. (\ref{eq:atmrange}) and (\ref{th13}), 
$\meff \sim 0.060$ eV.
The indicated values of $\meff$ are within 
the range of sensitivity of the next generation of
\betabeta-decay experiments.

  The expression for \meff, eq. (\ref{meffIH1}),
permits to relate the value of
$\sin^2 \alpha_{21}/2$ to the experimentally 
measured quantities \cite{BPP1,BGKP96}
$\meff$, $\dma$ and $\sin^22\theta_{\odot}$: 
\vspace{-0.1cm}
\begin{equation}
\sin^2  \frac{\alpha_{21}}{2}  \cong
\left( 1 - \frac{\meff^2}{|\dma| \cos^4\theta_{13}} \right) 
\frac{1}{\sin^2 2 \theta_\odot}~.
\end{equation}
%

\vspace{-0.2cm}
\noindent
A sufficiently accurate measurement of $\meff$
and of $|\dma|$ and  $\theta_\odot$,
could allow to get information 
about the value of $\alpha_{21}$,
provided the neutrino mass spectrum 
is of the $IH$ type.

\vspace{0.2cm} 
{\bf Three Quasi-Degenerate Neutrinos.}
 In this case 
$m_0 \equiv m_1 \cong  m_2 \cong m_3$,
$m^2_0 \gg |\dma|$, $m_0 \gtap 0.20$ eV. 
The mass $m_0$ effectively coincides with the 
$\bar{\nu}_e$ mass $m_{\bar{\nu}_e}$ 
measured in the $^{3}$H $\beta$-decay experiments:
$m_0 = m_{\bar{\nu}_e}$.   
Thus, $m_0 < 2.3 \eV$, or if we use a
conservative cosmological upper limit \cite{MTegmark},
$m_0 < 0.7$ eV.  The $QD$ 
$\nu$-mass spectrum is realized for 
values of $m_0$, which can be
measured in the $^3$H $\beta-$decay 
experiment KATRIN \cite{MainzKATRIN}.

 The effective Majorana mass $\meff$ is given by
\vspace{-0.2cm}
\begin{equation}
\meff \cong m_0 \left| (\cos^2 \theta_\odot
 + ~\sin^2 \theta_\odot 
 e^{i \alpha_{21}})~\cos^2\theta_{13} + 
 e^{i \alpha_{31}} \sin^2\theta_{13} \right| 
\label{meffQD0}
\end{equation}
\vspace{-0.6cm}
\begin{equation}
\cong m_0~\left| \cos^2 \theta_\odot + 
~\sin^2 \theta_\odot e^{i \alpha_{21}} \right|
= m_0~\sqrt{1 - \sin^22\theta_{\odot} 
\sin^2\frac{\alpha_{21}}{2}}.
\label{meffQD1}
\end{equation}
%

\vspace{-0.1cm}
\noindent Similarly to the case of $IH$ spectrum, one has:
\vspace{-0.1cm}
\begin{equation}
m_0~\left| \cos2\theta_\odot \right|~ 
\ltap \meff \ltap m_0~. 
\label{meffQD2}
\end{equation}
%

\vspace{-0.2cm}
\noindent For $\cos 2 \theta_\odot \sim 0.40$,
favored by the 
data, one finds a non-trivial lower limit
on $\meff$, $\meff \gtap 0.08$ eV.
For the 90\% C.L. allowed ranges of values 
of the parameters one has
$\meff \gtap 0.06$ eV. 
Using the conservative cosmological 
upper bound on $\sum_{j} m_{j}$ we get 
\begin{figure}[htb]
\vskip -0.5cm
\includegraphics[width=7.5cm,height=5.5cm]{KL2meff90SK3nus130.epsi}
\includegraphics[width=7.5cm,height=5.5cm]{KL2meff90SK3nus1304.epsi}
\vskip -0.3cm
\caption{The value of $\meff$ 
as function of $min(m_j)$
for $\sin^2\theta_{13}$=0.0;0.04 and 
90\% C.L. allowed ranges \cite{BCGPRKL2}
of $\dma$, $\dmsol$ and 
$\theta_{\odot}$ (updated version of Fig. 2 from \cite{PPSNO2bb}).
}
\label{Fig1}
\end{figure}
%
\noindent $\meff < 0.70$ eV.
Also in this case one can obtain, 
in principle, a direct information
on one $CPV$ phase from the measurement of $\meff$,
$m_0$ and $\sin^2 2 \theta_\odot$:
\vspace{-0.1cm}
\begin{equation}
\sin^2  \frac{\alpha_{21}}{2}  \cong 
\left( 1 - \frac{\meff^2}{m^2_0} \right) 
\frac{1}{\sin^2 2 \theta_\odot}.
\end{equation}

\vspace{-0.2cm}
 The specific features of the 
predictions for $\meff$ in the cases of 
the three types of neutrino mass 
spectrum discussed above are evident 
in Fig. 2, where the
dependence of $\meff$ on $min(m_j)$ for the 
LMA solution is shown. If
the spectrum is with normal hierarchy, 
$\meff$ can lie anywhere
between 0 and the presently existing upper limits, 
eqs. (\ref{76Ge00}) and (\ref{NEMO3CUOR}).
This conclusion does not change even 
under the most favorable
conditions for the determination of $\meff$,
namely, even when $|\dma|$, $\dmsol$,
$\theta_{\odot}$ and $\theta_{13}$ are known
with negligible uncertainty. 
If the results in \cite{Klap04} implying
$\meff = (0.1 - 0.9)~{\rm eV}$ are confirmed,
this would mean, in particular, that
the neutrino mass spectrum is of the $QD$ type.

\vspace{-0.6cm}
\section{{\large Implications of Measuring 
$\meff \neq 0$ }}

\vspace{-0.3cm}
  If the \betabeta-decay of a given nucleus 
will be observed, it would be possible to 
determine the value of $\meff$ from the
measurement of the associated half-life of the decay.
This would require the knowledge of the nuclear
matrix element of the process. 

\vspace{0.2cm}
{\bf On the NME Uncertainties.} At present there
exist large uncertainties in the calculation of
the \betabeta-decay nuclear matrix elements
(see, e.g., \cite{ElliotVogel02}).
This is reflected, in particular, in the
factor of $\sim 3$ uncertainty
in the upper limit on $\meff$, which is
extracted from the experimental
lower limits on the \betabeta-decay 
half-life of $^{76}$Ge. 
Recently, encouraging results 
in what regards the problem of the calculation
of the nuclear matrix elements have been
obtained in \cite{FesSimVogel03}. The observation of
a \betabeta-decay of one nucleus is likely to lead
to the searches and eventually to observation
of the decay of other nuclei.
One can expect that such a progress, in particular,
will help to solve the problem 
of the sufficiently precise calculation
of the nuclear matrix elements for the \betabeta-decay
\cite{NMEBiPet04}.

\vspace{0.2cm}
{\bf Constraining the Lightest Neutrino Mass.}
As Fig. 2 indicates, a measurement
of $\meff \gtap 0.01$ eV would either
i) determine a relatively narrow 
interval of possible values 
of the lightest 
$\nu$-mass $min(m_j) \equiv \mmin$, or 
ii) would establish an upper limit on 
$\mmin$. If an upper limit on 
$\meff$ is experimentally 
obtained below 0.01 eV,
this would lead to a significant upper limit on $\mmin$ 
and would imply $\dma >0$ 
for massive Majorana neutrinos.

 A measurement  of $\meff = (\meff)_{exp}
\gtap 0.02$ eV if $\dma \equiv \Delta m^2_{31} > 0$,
and of $\meff = (\meff)_{exp}
\gtap \sqrt{|\dma|}\cos^2\theta_{13}$
in the case of $\dma \equiv \Delta m^2_{32} < 0$,
for instance, would imply that 
$\mmin \gtap 0.02$ eV and 
$\mmin \gtap 0.05$ eV, respectively, and thus a 
$\nu$-mass spectrum with {\it partial hierarchy}
or of  $QD$ type \cite{BPP1}. 
The mass $\mmin$
will be constrained to lie in a rather 
narrow interval \cite{PPW} (Fig. 2).
If the measured value of $\meff$, $(\meff)_{exp}$, lies 
between the $min(\meff)$ and $max(\meff)$, predicted 
in the case of $IH$ spectrum, 
\vspace{-0.2cm}
\begin{equation}
\meff_{\pm} = 
\left |\sqrt{|\dma| - \dmsol}
\cos^2 \theta_\odot \pm \sqrt{|\dma|}
\sin^2 \theta_\odot \right | \cos^2\theta, 
\label{0m1ih}
\end{equation}
%

\vspace{-0.3cm}
\noindent 
$\mmin$ would be limited from above, but 
$min(\mmin)$=0 (Fig. 2).
If $(\meff)_{exp} < (\meff)_{max}$, where, 
e.g., in the case of $QD$ spectrum 
$(\meff)_{max} \cong \mmin \cong m_{\bar{\nu}_e}$,
this would imply that at least one of the two 
$CPV$ phases is different from zero :
$\alpha_{21}\neq 0$ and/or $\alpha_{31} \neq 0$
\footnote{In general, the knowledge of the value
of $\meff$ alone will not allow to 
distinguish the case
of $CP$-conservation 
from that of $CP$-violation.
}.

  A measured value of $m_{\bar{\nu}_e}$,
$(m_{\bar{\nu}_e})_{exp} \gtap 0.20$ eV, satisfying 
$(m_{\bar{\nu}_e})_{exp} > (\mmin)_{max}$,
where $(\mmin)_{max}$ is determined 
from the upper limit on $\meff$
in the case the \betabeta-decay is not 
observed, might imply 
that the massive neutrinos are Dirac particles.
If \betabeta-decay has been observed and $\meff$
measured, the inequality
$(m_{\bar{\nu}_e})_{exp} > (\mmin)_{max}$, with
$(\mmin)_{max}$ determined from 
the measured value of $\meff$,
would lead to the conclusion that
there exist contribution(s) to
the \betabeta-decay rate other than 
due to the light Majorana neutrino exchange
that partially cancels the
contribution from the 
Majorana neutrino exchange.

\vspace{0.2cm}
{\bf Determining the Type of Neutrino Mass Spectrum.}
  The existence of significant 
lower bounds on $\meff$
in the cases of $IH$ and 
$QD$ spectra \cite{PPSNO2bb},
which lie either partially ($IH$ spectrum) or completely
($QD$ spectrum) within the range of sensitivity of 
next generation of \betabeta-decay experiments,
is one of the most important features of
the predictions of $\meff$. 
These minimal values are given, 
up to small corrections, by
$\sqrt{|\dma|} \cos2\theta_{\odot}$ and
$m_0 \cos2\theta_{\odot}$. 
According to the
combined analysis of the solar and reactor 
neutrino data \cite{BCGPRKL2}
i) $\cos2\theta_{\odot}$ = 0
is excluded at $\sim$6$\sigma$,
ii) the best fit value of 
$\cos2\theta_{\odot}$ is 
$\cos2\theta_{\odot}$= 0.44, and
iii) at 95\% C.L. one has
for $\sin^2\theta_{13}$= 0~(0.02),
$\cos{2\theta_{\odot}} \gtap$ 0.28~(0.30).
The quoted results on $\cos{2\theta_{\odot}}$
together with the range of 
possible values of 
$|\dma|$ and $m_0$ 
\cite{SKYoichi04,Maltoni4nu,MainzKATRIN,MTegmark},
lead to the conclusion about the existence
of significant and robust lower 
bounds on $\meff$ in the cases of 
$IH$ and $QD$ spectrum \cite{PPSNO2bb,Carlosbb03}.
At the same time, as Fig. 2 indicates,
$\meff$ does not exceed $\sim 0.006$ eV
for $NH$ spectrum. 
This implies that 
$max(\meff)$ in the case of $NH$ spectrum 
is considerably smaller than $min(\meff)$
for the $IH$ and $QD$ spectrum.
This opens the 
possibility of obtaining 
information about the type of 
$\nu$-mass spectrum from a measurement of 
$\meff \neq 0$.
In particular, a positive result 
in the future generation
of \betabeta-decay experiments 
showing that $\meff > 0.02$ eV 
would imply that the $NH$ spectrum is 
strongly disfavored (if not excluded).
The uncertainty in the relevant NME
and prospective experimental errors 
in  the values of the oscillation parameters
and in $\meff$
can weaken, but do not invalidate, 
these results 
(see, e.g., ref. \cite{PPRSNO2bb}).

\vspace{0.2cm}
{\bf Constraining the Majorana $CPV$ Phases.}
The possibility of establishing 
$CP$-violation 
due to Majorana $CPV$ 
phases has been studied in \cite{PPW} and in 
greater detail in \cite{PPR1}.
It was found that it is very challenging
\footnote{Pessimistic conclusion 
about the prospects to establish  $CP$-violation 
due to Majorana $CPV$ phases
from a measurement of $\meff$ and, e.g., of 
$m_0$, was reached in \cite{BargerCP}.}:
it requires quite accurate measurements 
of $\meff$ and of $m_0$, 
and holds only for a limited range of 
values of the relevant parameters.
For $IH$ and $QD$ spectra, which are 
of interest, the ``just CP-violation'' 
region~\cite{BPP1}
- an experimental point in this region
would signal unambiguously CP-violation
associated with Majorana neutrinos,
is larger for smaller values of 
$\cos 2 \theta_\odot$. 
More specifically, 
proving that $CP$-violation associated with
Majorana neutrinos takes place
requires, in particular, a relative 
experimental error on the measured value of 
$\meff$ smaller than $\sim15$\%,
a ``theoretical uncertainty'' in the value of
$\meff$ due to an imprecise knowledge of the 
corresponding $NME$
smaller than a factor of 2, a value of 
$\tan^2\theta_{\odot} \gtap 0.55$,
and values of the relevant Majorana
$CPV$ phases 
typically within the ranges 
of $\sim (\pi/2 - 3\pi/4)$ and
$\sim (5\pi/4 - 3\pi/2)$. 

\vspace{-0.60cm}
\section{{\large Conclusions}} 
\vspace{-0.3cm}

 Future $\betabeta-$decay experiments 
have a remarkable physics potential. They can establish 
the Majorana nature of the neutrinos with definite mass
$\nu_j$. If the latter are Majorana particles,
the $\betabeta-$decay experiments can provide 
information on the 
type of the neutrino mass spectrum and 
on the absolute scale of neutrino masses.
They can also provide unique information on the
Majorana $CP$-violation phases present in the 
PMNS neutrino mixing matrix. 
The knowledge of the 
values of the relevant $\betabeta-$decay nuclear 
matrix elements with a sufficiently small uncertainty
is crucial for obtaining quantitative information on
the neutrino mass and mixing parameters 
from a measurement of
$\betabeta-$decay half-life.

\vspace{-0.50cm}

\end{document}